\documentclass[aps,prd,nofootinbib,amsmath,amssymb,superscriptaddress,twocolumn,10pt]{revtex4}

\pdfoutput=1

\usepackage{graphicx}
\usepackage{dcolumn}
\usepackage{bm}
\usepackage{amssymb}
\usepackage{latexsym}
\usepackage{booktabs}
\usepackage{amsmath}
\usepackage{multirow}
\usepackage[colorlinks=true, linkcolor=red, citecolor=blue]{hyperref}
\usepackage{mathrsfs}

\newcommand{\be}{\begin{equation}}
\newcommand{\ee}{\end{equation}}
\newcommand{\bq}{\begin{eqnarray}}
\newcommand{\eq}{\end{eqnarray}}

\begin{document}

\title{Constraining neutrino mass in dynamical dark energy cosmologies with the logarithm parametrization and the oscillating parametrization}

\author{Tian-Ying Yao}
\affiliation{School of Sciences, Xi'an Technological University, Xi'an 710021, China}
\author{Rui-Yun Guo\footnote{Corresponding author}}
\email{guoruiyun@xatu.edu.cn}
\affiliation{School of Sciences, Xi'an Technological University, Xi'an 710021, China}
\author{Xin-Yue Zhao}
\affiliation{School of Sciences, Xi'an Technological University, Xi'an 710021, China}

\begin{abstract}
	
We constrain two dynamical dark energy models that are parametrized by the logarithm form of $w(z)=w_{0}+w_{1}\left(\frac{\ln (2+z)}{1+z}-\ln 2\right)$ and the oscillating form of $w(z)=w_{0}+w_{1}\left(\frac{\sin(1+z)}{1+z}-\sin(1)\right)$. Comparing with the Chevallier-Polarski-Linder (CPL) model, the two parametrizations for dark energy can explore the whole evolution history of the universe properly. Using the current mainstream observational data including the cosmic microwave background data and the baryon acoustic oscillation data as well as the type Ia supernovae data, we perform the $\chi^2$ statistic analysis to global fit these models, finding that the logarithm parametrization and the oscillating parametrization are almost as well as the CPL scenario in fitting these data. We make a comparison for the impacts of the dynamical dark energy on the cosmological constraints on the total mass of active neutrinos. We find that the logarithm parametrization and the oscillating parametrization can increase the fitting values of $\sum m_{\nu}$. Looser constraints on $\sum m_{\nu}$ are obtained in the logarithm and oscillating models than those derived in the CPL model. Consideration of the possible mass ordering of neutrinos reveals that the most stringent constraint on $\sum m_{\nu}$ appears in the degenerate hierarchy case.

\end{abstract}

\maketitle

\section{Introduction}\label{sec:1}

The fact that neutrinos have masses~\cite{Lesgourgues:2006nd,Agashe:2014kda} has drawn significant attention from physicists. The squared mass difference between different neutrino species have been measured, i.e., $\Delta m_{21}^{2}\simeq 7.5\times 10^{-5}$ eV$^{2}$ in solar and reactor experiments, and $|\Delta m_{31}^{2}|\simeq 2.5\times 10^{-3}$ eV$^{2}$ in atmospheric and accelerator beam experiments~\cite{Agashe:2014kda}. The possible mass hierarchies of neutrinos are $m_{1}<m_{2}\ll m_{3}$ and $m_{3}\ll m_{1}< m_{2}$, which are called the normal hierarchy (NH) and the inverted hierarchy (IH). When the mass splittings between different neutrino species are neglected, we treat the case as the degenerate hierarchy (DH) with $m_{1}=m_{2}=m_{3}$. 

Some famous particle physics experiments, such as tritium beta decay experiments~\cite{KATRIN:2001ttj,Kraus:2004zw,Otten:2008zz,Wolf:2008hf} and neutrinoless double beta decay (0$\nu \beta \beta$) experiments~\cite{KlapdorKleingrothaus:2002ip,KlapdorKleingrothaus:2004wj}, have been designed to measure the absolute masses of neutrinos. Recently, the Karlsruhe Tritium Neutrino (KATRIN) experiment provided an upper limit of $1.1$ eV on the neutrino-mass scale at $2 \sigma$ confidence level (C.L.)~\cite{KATRIN:2021fgc}. However, cosmological observations are considered to be a more promising approach to measure the total neutrino mass $\sum m_{\nu}$. Massive neutrinos can leave rich imprints on the cosmic microwave background (CMB) anisotropies and the large-scale structure (LSS) formation in the evolution of the universe. Thus, the total neutrino mass $\sum m_{\nu}$ is likely to be measured from these available cosmological observations.

In the standard $\Lambda$ cold dark matter ($\Lambda$CDM) model with the equation-of-state parameter of dark energy $w=-1$, the Planck Collaboration gave $\sum m_{\nu}<0.26$ eV ($2 \sigma$)~\cite{Planck:2018vyg} from the full Planck TT, TE, EE power spectra data, assuming the NH case with the minimal mass $\sum m_{\nu}=0.06$ eV ($2 \sigma$). Adding the Planck CMB lensing data slightly tightens the constraints to $\sum m_{\nu}<0.24$ eV ($2 \sigma$). When the baryon acoustic oscillations (BAO) data are considered on the basis of the Planck data, the neutrino mass constraint is significantly tightened to $\sum m_{\nu}<0.12$ eV ($2 \sigma$). Further adding the type Ia supernovae (SNe) data marginally lowers the bound to $\sum m_{\nu}<0.11$ eV ($2 \sigma$), which put pressure on the inverted mass hierarchy with  $\sum m_{\nu} \ge 0.10$ eV.  

The impacts of dynamical dark energy on the total neutrino mass have been investigated in past studies~\cite{Zhao:2016ecj,Zhang:2015uhk,Li:2012vn,Wang:2012uf,Guo:2018gyo,Zhang:2014nta,Zhang:2015rha,Wang:2018lun,Wang:2016tsz,Yang:2017amu,Huang:2015wrx,Zhang:2020mox,Sharma:2022ifr,RoyChoudhury:2019hls,Geng:2015haa,Chen:2015oga,Vagnozzi:2018jhn,Loureiro:2018pdz,Riess:2019cxk,Giusarma:2016phn,Vagnozzi:2017ovm,Giusarma:2018jei,Tanseri:2022zfe,Khalifeh:2021ree,Yang:2019nnl,Lee:2020uag,Xu:2020fyg}. In the simplest dynamical dark energy model with $w=Constant$ (abbreviated as $w$CDM model), the fitting results of $\sum m_{\nu}$ are $\sum m_{\nu, \rm NH}<0.195$ eV ($2 \sigma$) and $\sum m_{\nu, \rm IH}<0.220$ eV ($2 \sigma$)~\cite{Zhang:2020mox}, using the full Planck TT, TE, EE power spectra data and the BAO data as well as the SNe data. From the same data combination, $\sum m_{\nu, \rm NH}<0.129$ eV ($2 \sigma$) and $\sum m_{\nu, \rm IH}<0.163$ eV ($2 \sigma$)~\cite{Zhang:2020mox} in the holographic dark energy (HDE) model~\cite{Li:2004rb,Huang:2004ai,Zhang:2014ija,Wang:2016och,Wang:2013zca,Cui:2015oda,He:2016rvp,Xu:2016grp}. The constraint results of $\sum m_{\nu}$ are different from those in the standard $\Lambda$CDM model because of impacts of dark energy properties in these cosmological models.

In addition to the $w$CDM model and the HDE model, the constraints on $\sum m_{\nu}$ are investigated in the CPL model~\cite{Chevallier:2000qy,Linder:2002et} with $w(z)=w_{0}+w_{1}\frac{z}{1+z}$ (where $w_{0}$ and $w_{1}$ are two free parameters). Over the years, the CPL parametrization have been widely used and explored extensively. In the model, $\sum m_{\nu, \rm NH}<0.290$ eV ($2 \sigma$) and $\sum m_{\nu, \rm IH}<0.305$ eV ($2 \sigma$)~\cite{Zhang:2020mox} are obtained by using the full Planck TT, TE, EE power spectra data combined the BAO data with the SNe data. The upper limit values of $\sum m_{\nu}$ are larger than those in the $w$CDM model and the HDE model, confirming that the constraint results of $\sum m_{\nu}$ can be changed as the different parametrization forms of $w$. The CPL model has a drawback that it only explores the past expansion history, but cannot describe the future evolution (Owing to that $|w(z)|$ grows increasingly and finally encounters divergency as $z\rightarrow -1$). Thus the CPL parametrization does not genuinely cover the scalar field models as well as other theoretical models. Such a problem makes the fitting results of $\sum m_{\nu}$ untenable in the CPL model. 

To investigate the impacts of two-parametrization dynamical dark energy on the total neutrino mass $\sum m_{\nu}$ physically, we focus on two special dynamical dark energy models that are proposed in Ref.~\cite{Ma:2011nc} with the logarithm parametrization and the oscillating parametrization. They are indicated to be more favored than the CPL model by the observational data~\cite{Ma:2011nc}. For convenience, the two models are called the Log model and the Sin model, hereafter. For the Log model, $w(z)=w_{0}+w_{1}\left(\frac{\ln (2+z)}{1+z}-\ln 2\right)$. Thus we have 
\begin{eqnarray}
{w(z)} = \left\{\begin{array}{ll} w_{0},\ \ \ \ \ \ \ \ \ \ &$for$\
z=0,\\\\
w_{0}-w_{1}\ln2,&$for$\ z\rightarrow+\infty,\\\\
w_{0}+w_{1}(1-\ln2),&$for$\ z\rightarrow-1.
\end{array}\right.
\end{eqnarray}
Such a parametrization can exhibit well-behaved feature for the dynamical evolution of dark energy. $w(z)=w_{0}$ (the value of $w(z)$ in current cosmology) at $z=0$. When $z\rightarrow+\infty$ (i.e., at high redshifts) and $z\rightarrow-1$ (i.e., at negative redshifts), a finite value for $w(z)$ can be ensured, successfully avoiding the future divergency problem in the CPL model. 

For the Sin model that considers the possible oscillating feature during the evolution of dark energy,  $w(z)=w_{0}+w_{1}\left(\frac{\sin(1+z)}{1+z}-\sin(1)\right)$. Comparing with the logarithm parametrization, the change is that the logarithm function is replaced with a sine function. In this situation, 
\begin{eqnarray}
{w(z)} = \left\{\begin{array}{ll} w_{0},\ \ \ \ \ \ \ \ \ \ &$for$\
z=0,\\\\
w_{0}-w_{1}\sin(1),&$for$\ z\rightarrow+\infty,\\\\
w_{0}+w_{1}(1-\sin(1)),&$for$\ z\rightarrow-1.
\end{array}\right.
\end{eqnarray}
When $z=0$, $w(z)=w_{0}$, that still corresponds to the $w$CDM model with a free parameter $w_{0}$. Since $\sin(1)\approx \ln2$, the two parametrizations are almost identical at low redshifts and can describe the same behavior of dynamical dark energy. The difference is that the oscillating parametrization exhibits oscillating feature from a long term point of view. Similarly, when $z\rightarrow+\infty$ and $z\rightarrow-1$, the two parametrizations also roughly coincide in the limiting cases and do not encounter divergency of $w(z)$ during the whole evolution of the universe.

The reasons for choosing the two parametrizations are in this work: (i) They can exhibit well-behaved features for the dynamical evolution of dark energy. (ii) They are indicated to be more favored than the CPL model by the observational data~\cite{Ma:2011nc}. (iii) They can successfully avoid the future divergency problem in the CPL parametrization, and help probe the dynamics of dark energy in the whole evolutionary history. For more relevant studies for the two parametrizations, please refer to the references~\cite{Pan:2019brc,DiValentino:2020evt,Perkovic:2020mph,Pacif:2020hai,Cardenas:2020srs,Ren:2021tfi,Rezaei:2021qwd,Rezaei:2022bkb,Wang:2022jpo,Yang:2022kho}. In fact, there are also some other two-parameter forms of $w(z)$ that can describe the dynamical evolution of dark energy, such as the Jassal-Bagla-Padmanabhan parametrization~\cite{Jassal:2005qc} and the Barboza-Alcaniz parametrization~\cite{Barboza:2008rh}. They both deserve a detailed discussion in future research. These previous researches have indicated that the nature of dark energy can change the total neutrino mass. Aside from the theory of dark energy, another popular explanation for cosmic acceleration is a modification of Einstein’s general relativity, i.e., modified gravity (MG)~\cite{Sahni:1998at,Nicolis:2008in,Linder:2010py,Corda:2009re,Corda:2011ri}. They both can provide the negative energy pressure to realize cosmic acceleration. Thus, it is also a significant task to explore possible impact of the modified gravity on cosmological constraints on the neutrino mass.

In our present work, we revisit the constraints on dynamical dark energy that is parametrized by the logarithm form and the oscillating form, by using latest mainstream observational data. Impacts of the logarithm and oscillating parametrizations of $w(z)$ on the fitting results of $\sum m_{\nu}$ are investigated for the first time. Meanwhile, we also consider the three mass hierarchies of neutrinos (NH, IH, and DH), and analyze the effect of different mass hierarchies of neutrinos on $\sum m_{\nu}$. In addition, in order to better match the current observational result of $w=-1$, we assume the case of $w_{0}=-1$ in the logarithm parametrization and the oscillating parametrization. The forms of $w(z)$ in these models are modified to be $w(z)=-1+w_{1}\left(\frac{\ln (2+z)}{1+z}-\ln 2\right)$ and $w(z)=-1+w_{1}\left(\frac{\sin(1+z)}{1+z}-\sin(1)\right)$ with a free parameter $w_{1}$. They still describe the logarithm feature and the oscillating feature during the evolution of dynamical dark energy, respectively. We call them the MLog model and the MSin model. We also investigate the constraints on the one-parameter dark energy by using the same mainstream observational data. We want to probe how one-parameter logarithm and oscillating parametrizations of $w(z)$ influence on the fitting results of $\sum m_{\nu}$. 

This paper is organized as follows. In Sect.~\ref{sec:2}, we provide a brief description of the data and method used in our work. In Sect.~\ref{sec:3}, we show the constraint results of different dynamical dark energy models and discuss the physical meaning behind these results. At last, we make some important conclusions in Sect.~\ref{sec:4}.

\section{Data and method}\label{sec:2}

Throughout this paper, we only employ the data combination of the CMB data, the BAO data, and the SNe data, which is abbreviated as the CMB+BAO+SNe data. The usage of the data combination facilitates to make a comparison with the results derived from Refs.~\cite{Planck:2018vyg,Zhang:2020mox,RoyChoudhury:2019hls}, in which this typical data combination has also been used to constrain cosmological models. For the CMB data, we use the Planck 2018 temperature and polarization power spectra data at the whole multipole ranges, together with the CMB latest lensing power spectrum data~\cite{Planck:2018vyg}. For the BAO data, we use the 6dFGS and SDSS-MGS measurements of $D_{\rm V}/r_{\rm drag}$~\cite{Beutler:2011hx,Ross:2014qpa} plus the final DR12 anisotropic BAO measurements~\cite{BOSS:2016wmc}. For the SNe data, we use the “Pantheon” sample~\cite{Pan-STARRS1:2017jku}, which contains 1048 supernovae covering the redshift range of $0.01<z<2.3$.

In our present work, we assume a spatially flat universe with its Friedmann equation
\begin{equation}\label{2.1}
H(z)^{2}=\frac{8\pi G}{3} [\rho_{\rm r0}(1+z)^{4}+\rho_{\rm m0}(1+z)^{3}+\rho_{\rm de}(z)],
\end{equation} 
where $H(z)$ is the Hubble expansion rate, $\rho_{\rm r0}$ and $\rho_{\rm m0}$ are the radiation density and matter density in current cosmology. $\rho_{\rm de}(z)$ refers to the energy density of dark energy, and can be written as
\begin{equation}\label{2.2}
\rho_{\rm de}(z)=\rho_{\rm de0} \exp \left\{3 \int^{z}_{0} \frac{dz^{\prime}}{(1+z^{\prime})} [1+w(z^{\prime})]\right\},
\end{equation}
where $\rho_{\rm de0}$ is the current value of dark energy density. The Hubble expansion rate $H(z)$ is affected by dynamical evolutuon of dark energy.

For the dynamical dark energy models with the CPL parametrization, logarithm, and oscillating parametrizations, they all have eight free parameters, i.e., the present baryons density $\omega_{\rm b}\equiv \Omega_{\rm b}h^{2}$, the present cold dark matter density $\omega_{\rm c}\equiv \Omega_{\rm c}h^{2}$, an approximation to the angular diameter distance of the sound horizon at the decoupling epoch $\theta_{\rm MC}$, the reionization optical depth $\tau$, the amplitude of the primordial scalar power spectrum $A_{\rm s}$ at $k=0.05$ Mpc$^{-1}$, the primordial scalar spectral index $n_{\rm s}$, and the model parameters $w_{0}$ and $w_{1}$. The priors of these parameters are shown explicitly in the Table~\ref{table}. When $w_{0}=-1$ is fixed, there are seven free parameters in the MCPL, MLog, and MSin models. When the influence from total mass $\sum m_{\nu}$ is not considered in these dynamical dark energy models, we uniformly assume $\sum m_{\nu}=0.06$ eV including two massless and one massive neutrino species.

\begin{table}[ht!]\tiny
	\caption{Priors on the free parameters for the two-parametrization dark energy models.}
	\label{table}
	\small
	\setlength\tabcolsep{2.8pt}
	\renewcommand{\arraystretch}{1.5}
	\centering
	\begin{tabular}{cccccccccccccccccc}
		\\
		\hline\hline
		Parameter & Prior   \\
		\cline{1-2}
		
		$\Omega_{\rm b} h^2$             
		& $[0.005, 0.100]$
		\\
		
		$\Omega_{\rm c} h^2$           
		& $[0.001, 0.990]$
		\\
		
		$100\theta_{\rm MC}$
		& $[0.5, 10.0]$
		\\
		
		$\tau$
		& $[0.01, 0.80]$
		\\
		
		${\rm{ln}}(10^{10}A_{\rm s})$
		& $[2, 4]$
		\\
		
		$n_{\rm s}$
		& $[0.8, 1.2]$
		\\
		
		\hline
		
		$w_{0}$
		& $[-3.0, -0.01]$
		\\
		
		$w_{1}$
		& $[-4, 9]$
		\\

		\hline\hline
	\end{tabular}
\end{table}

We consider the case that $\sum m_{\nu}$ serves as a free parameter with different hierarchies of neutrino mass. The neutrino$\_$hierarchy parameter in  the {\tt camb} Boltzmann code~\cite{Lewis:1999bs} can be set to normal or inverted, so that we adopt a two-eigenstate model that is a good approximation to the known mass splittings, then determining the total neutrino mass. For the NH, IH, and DH cases, the priors of $\sum m_{\nu}$ are $[0.06, 3.00]$ eV, $[0.10, 3.00]$ eV, and $[0.00, 3.00]$ eV. Correspondingly, the neutrino mass spectrum is described as 
\begin{equation*}
(m_{1},m_{2},m_{3})=(m_{1},\sqrt{m_{1}^{2}+\Delta m_{21}^{2}},\sqrt{m_{1}^{2}+|\Delta m_{31}^{2}|})
\end{equation*} with a free parameter $m_{1}$ for the NH case,
\begin{equation*}
(m_{1},m_{2},m_{3})=(\sqrt{m_{3}^{2}+|\Delta m_{31}^{2}|},\sqrt{m_{3}^{2}+|\Delta m_{31}^{2}|+\Delta m_{21}^{2}},m_{3})
\end{equation*} with a free parameter $m_{3}$ for the IH case, and
\begin{equation*}
m_{1}=m_{2}=m_{3}=m
\end{equation*} with a free parameter $m$ for the DH case.

In order to check the consistency between dynamical dark energy models and the CMB+BAO+SNe data, we employ the $\chi^2$ statistic~\cite{Guo:2021rrz,Feng:2021ipq,Guo:2018ans} to do the cosmological fits. A model with a lower value of $\chi^2$ is more favored by the CMB+BAO+SNe data combination. Our constraint results are derived by modifying the August 2017 version of the {\tt camb} Boltzmann code~\cite{Lewis:1999bs} and the July 2018 version of {\tt CosmoMC}~\cite{Lewis:2013hha}. For the calculation methods of the cosmological perturbations in these models, we adopt the default settings of the publicly available {\tt CosmoMC} package~\cite{Lewis:2013hha}, following the Planck collaboration~\cite{Planck:2018vyg}. 

\section{results and discussions}\label{sec:3}

We constrain the sum of the neutrino mass $\sum m_{\nu}$ in these dynamical dark energy models by using the CMB+BAO+SNe data. In the following discussion, we will present the fitting results with the $\pm1\sigma$ errors of cosmological parameters. But for the constraints on $\sum m_{\nu}$, we only provide the  $2\sigma$ upper limit. Meanwhile, we also list the values of $\chi^{2}_{\rm min}$ for different dark energy models.

\subsection{Comparison of dynamical dark energy models}\label{sec:3.1}

We constrain the models parameterized by $w(z)=w_{0}+w_{1}\frac{z}{1+z}$, $w(z)=w_{0}+w_{1}\left(\frac{\ln (2+z)}{1+z}-\ln 2\right)$ and $w(z)=w_{0}+w_{1}\left(\frac{\sin(1+z)}{1+z}-\sin(1)\right)$. The fitting results are listed in Table~\ref{tab1}. We find that the current CMB+BAO+SNe data favor the constraint results of $w_{0}=-1$ and $w_{1}=0$ in the three models. For the CPL model, we obtain $\Omega_{\rm m}=0.3059\pm0.0077$ and $H_0=68.37\pm0.83$ km/s/Mpc, with $\chi^{2}_{\rm min}=3821.214$. For the Log model, we have $\Omega_{\rm m}=0.3060\pm0.0075$ and $H_0=68.37\pm0.81$ km/s/Mpc, with $\chi^{2}_{\rm min}=3821.150$. For the Sin model, we have $\Omega_{\rm m}=0.3056\pm0.0077$ and $H_0=68.41\pm0.83$ km/s/Mpc, with $\chi^{2}_{\rm min}=3821.164$. The fitting values of $\Omega_{\rm m}$ and $H_0$ are similar for the three models. According to the $\chi^{2}_{\rm min}$ values, the models provide a similar fit to the CMB+BAO+SNe data. However, compared with $\chi^{2}_{\rm min}=3824.922$ in the base $\Lambda$CDM model~\cite{Guo:2021rrz}, the $\chi^{2}_{\rm min}$ values in these models are decreased by more than $3$ (corresponding to the relative value of the Akaike information criterion $\Delta {\rm AIC} <1$), thus we say that the three models are favored by the current observations.

\begin{table*}[ht!]\tiny
	\caption{The fitting values for the six dynamical dark energy models} 
	\label{tab1}
	\small
	\setlength\tabcolsep{2.8pt}
	\renewcommand{\arraystretch}{1.5}
	\centering
	\begin{tabular}{cccccccccccccccccc}
		\\
		\hline\hline
		Parameter & CPL  & Log  & Sin &   MCPL   &MLog  &MSin \\
		\cline{1-7}
		
		$w_{0}$             &$-0.968\pm0.079$
		&$-0.968^{+0.065}_{-0.072}$
		&$-0.973^{+0.059}_{-0.058}$
		&$-1$
		&$-1$
		&$-1$
		\\

		$w_{1}$           &$-0.24^{+0.33}_{-0.27}$
		&$0.93^{+0.79}_{-1.11}$
		&$0.36^{+0.28}_{-0.40}$
		&$-0.12^{+0.13}_{-0.11}$
		&$0.52^{+0.39}_{-0.48}$
		&$0.22^{+0.16}_{-0.21}$
		\\
		
		$\Omega_{\rm m}$
		& $0.3059\pm0.0077$
		& $0.3060\pm0.0075$
		& $0.3056\pm0.0077$
		& $0.3048^{+0.0070}_{-0.0071}$
		& $0.3045^{+0.0069}_{-0.0068}$
		& $0.3044\pm0.0068$
		\\
		
		$H_0$ [km/s/Mpc]                    &$68.37\pm0.83$
		&$68.37\pm0.81$
		&$68.41\pm0.83$
		&$68.47\pm0.76$
		&$68.53\pm0.73$
		&$68.55\pm0.73$
		\\
		
		$\sigma_8$                 &$0.822\pm0.011$
		&$0.822\pm0.011$
		&$0.823\pm0.011$
		&$0.822\pm0.011$
		&$0.823\pm0.011$
		&$0.824\pm0.011$
		\\
		
		\hline
		$\chi^{2}_{\rm min}$    &$3821.214$
		&$3821.150$
		&$3821.164$
		&$3821.310$
		&$3821.288$
		&$3821.290$
		\\
		\hline\hline
	\end{tabular}
\end{table*}

\begin{figure*}[ht!]
	\begin{center}
		\includegraphics[width=12cm]{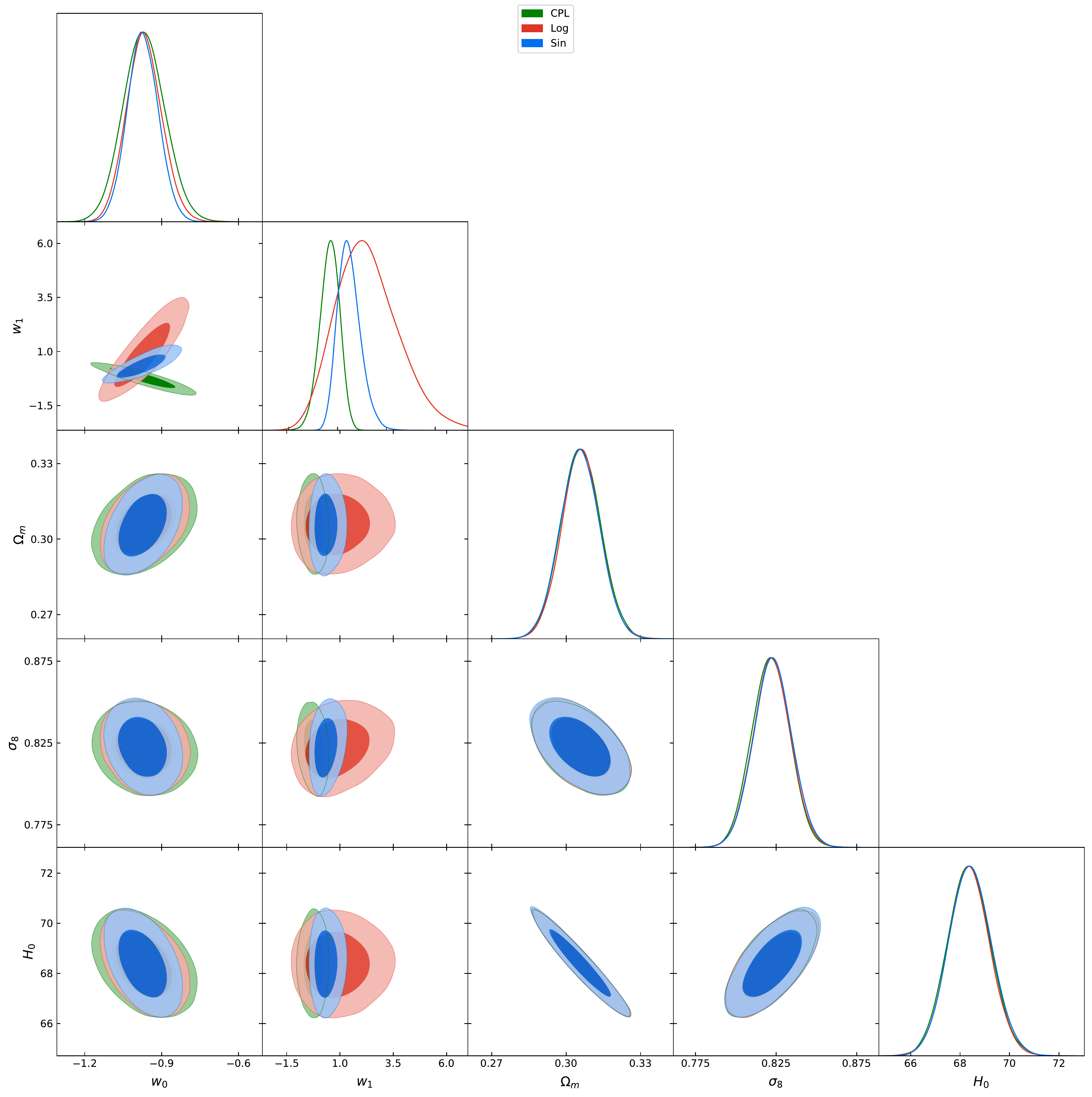}
	\end{center}
	\caption{One-dimensional marginalized distributions and two-dimensional contours at $1\sigma$ and $2\sigma$ level for the CPL, Log, and Sin models.}
	\label{f1}
\end{figure*}

\begin{figure*}[ht!]
	\begin{center}
		\includegraphics[width=12cm]{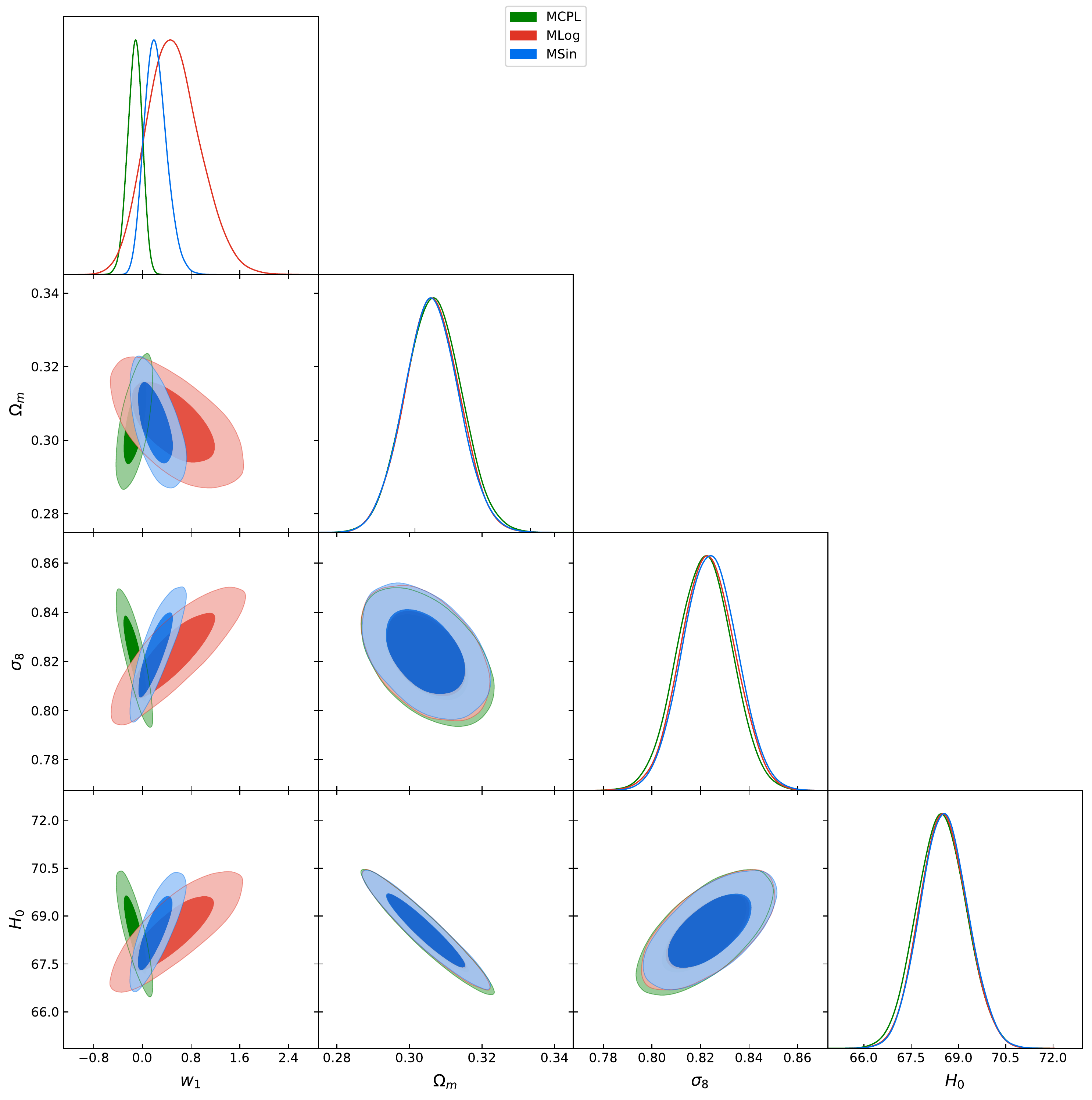}
	\end{center}
	\caption{One-dimensional marginalized distributions and two-dimensional contours at $1\sigma$ and $2\sigma$ level for the MCPL, MLog, and MSin models.}
	\label{f2}
\end{figure*}

As described in Sect.~\ref{sec:1}, when $w_{0}=-1$ is fixed in the above models, the form of $w(z)$ is modified with a free parameter $w_{1}$. The fitting results are also given in the last three columns of Table~\ref{tab1}. In the MCPL model, $w(z)=-1+w_{1}\frac{z}{1+z}$. In the MLog model, $w(z)=-1+w_{1}\left(\frac{\ln (2+z)}{1+z}-\ln 2\right)$. In the MSin model, $w(z)=-1+w_{1}\left(\frac{\sin(1+z)}{1+z}-\sin(1)\right)$. We obtain $w_{1}=-0.12^{+0.13}_{-0.11}$, $w_{1}=0.52^{+0.39}_{-0.48}$, and $w_{1}=0.22^{+0.16}_{-0.21}$, showing a slight deviation to $w_{1}=0$ in the MLog model and the MSin model. This is because $w_{1}$ is intrinsically correlated with $w_{0}$, as shown in Figure~\ref{f1} ($w_{1}$ is anticorrelated with $w_{0}$ in the CPL model, but the correlation between them is opposite in the Log model and the Sin model). When the value of $w_{0}$ is fixed to $-1$, the fitting value of $w_{1}$ will be changed to a certain extent. 

Furthermore, we focus on the $\chi^{2}_{\rm min}$ values for the three models. We obtain $\chi^{2}_{\rm min}=3821.310$ in the MCPL model, $\chi^{2}_{\rm min}=3821.288$ in the MLog model, and $\chi^{2}_{\rm min}=3821.290$ in the MSin model. Similarly, almost identical $\chi^{2}_{\rm min}$ values are presented in the three models. In Figures~\ref{f1} and \ref{f2}, we also provide the one-dimensional marginalized distributions and two-dimensional contours at $1\sigma$ and $2\sigma$ level for these dynamical dark energy models. The fitting results of the parameter $\Omega_{\rm m}$, $H_0$, and $\sigma_8$ hardly change in these models despite of $w(z)$ parametrized by different forms.

\subsection{Constraints on neutrino masses}

We investigate the constraints on total neutrino mass in these models. For the neutrino mass measurement, we consider the NH case, the IH case, and the DH case. The fitting results are listed in Tables~\ref{tab2}$-$~\ref{tab4}. In the CPL+$\sum m_{\nu}$ model, we obtain $\sum m_{\nu}<0.285$ eV for the NH case, $\sum m_{\nu}<0.304$ eV for the IH case, and $\sum m_{\nu}<0.254$ eV for the DH case (see Table~\ref{tab2}). In the Log+$\sum m_{\nu}$ model, we have $\sum m_{\nu}<0.302$ eV for the NH case, $\sum m_{\nu}<0.317$ eV for the IH case, and $\sum m_{\nu}<0.282$ eV for the DH case (see Table~\ref{tab3}), showing that much looser constraints are obtained than those in the CPL+$\sum m_{\nu}$ model. In the Sin+$\sum m_{\nu}$ model, the constraint results become $\sum m_{\nu}<0.327$ eV for the NH case, $\sum m_{\nu}<0.336$ eV for the IH case, and $\sum m_{\nu}<0.311$ eV for the DH case (see Table~\ref{tab4}), which are looser than those in the Log+$\sum m_{\nu}$ model. All the above fitting upper limits on $\sum m_{\nu}$ are larger than those obtained in the standard $\Lambda$CDM model (in the $\Lambda$CDM model, the constraint results are $\sum m_{\nu}<0.156$ eV for the NH case, $\sum m_{\nu}<0.184$ eV for the IH case, and $\sum m_{\nu}<0.121$ eV for the DH case~\cite{Zhang:2020mox,Jin:2022tdf}), indicating that the dynamical dark energy with the logarithm form and the oscillating form can affect significantly the fitting value of $\sum m_{\nu}$. 

\begin{table*}\tiny
	\caption{The fitting values for the CPL+$\sum m_{\nu}$ and MCPL+$\sum m_{\nu}$ models considered mass hierarchy cases of
		NH, IH, and DH.}
	\label{tab2}
	\small
	\setlength\tabcolsep{2.8pt}
	\renewcommand{\arraystretch}{1.5}
	\centering
	\begin{tabular}{cccccccccccc}
		\\
		\hline\hline &\multicolumn{3}{c}{CPL} &&\multicolumn{3}{c}{MCPL} \\
		\cline{2-4}\cline{6-8}
		Parameter  &NH &IH &DH &&NH &IH  &DH  \\ \hline
		
		$w_{0}$                & $-0.940^{+0.085}_{-0.095}$
		& $-0.929^{+0.083}_{-0.097}$
		& $-0.950^{+0.082}_{-0.092}$
		&& $-1$
		& $-1$
		& $-1$\\

		$w_{1}$       & $-0.49^{+0.46}_{-0.33}$
		&$-0.59^{+0.48}_{-0.32}$
		& $-0.39^{+0.47}_{-0.30}$
		&& $-0.24^{+0.18}_{-0.13}$
		& $-0.30^{+0.18}_{-0.14}$
		& $-0.17^{+0.19}_{-0.13}$\\
		
		$\sum m_{\nu}$ [eV]         &  $<0.285$
		&$<0.304$
		&  $<0.254$
		&&$<0.250$
		& $<0.276$
		& $<0.228$\\
		
		$\Omega_{\rm m}$       & $0.3094^{+0.0081}_{-0.0087}$
		& $0.3103^{+0.0081}_{-0.0082}$
		& $0.3077^{+0.0083}_{-0.0090}$
		&& $0.3069\pm0.0073$
		&$0.3078\pm0.0072$
		& $0.3058^{+0.0072}_{-0.0079}$\\
		
		$H_0$ [km/s/Mpc]        & $68.27\pm0.82$
		&$68 .27^{+0.83}_{-0.81}$
		& $68.32\pm0.84$
		&& $68.47\pm0.76$
		& $68 .49\pm0.75$
		& $68.45^{+0.77}_{-0.76}$\\

		$S_8$       &$0.825\pm0.012$
		&$0.823\pm0.012$
		&$0.827\pm0.012$
		&&$0.824\pm0.011$
		&$0.822\pm0.011$
		&$0.826\pm0.012$\\

		\hline
		$\chi^{2}_{\rm min}$    &$3822.102$
		&$3822.516$
		&$3821.168$
		&&$3822.144$
		&$3823.046$
		&$3821.112$
		\\	
		\hline\hline
	\end{tabular}
\end{table*}

\begin{table*}\tiny
	\caption{The fitting values for the Log+$\sum m_{\nu}$ and MLog+$\sum m_{\nu}$ models considered mass hierarchy cases of
		NH, IH, and DH.}
	\label{tab3}
	\small
	\setlength\tabcolsep{2.8pt}
	\renewcommand{\arraystretch}{1.5}
	\centering
	\begin{tabular}{cccccccccccc}
		\\
		\hline\hline &\multicolumn{3}{c}{Log} &&\multicolumn{3}{c}{MLog} \\
		\cline{2-4}\cline{6-8}
		Parameter  &NH &IH &DH &&NH &IH  &DH  \\ \hline
		
		$w_{0}$                & $-0.946^{+0.071}_{-0.080}$
		&$-0.938^{+0.073}_{-0.081}$
		&$-0.955^{+0.069}_{-0.079}$
		&& $-1$
		& $-1$
		& $-1$\\

		$w_{1}$       & $1 .90^{+1.00}_{-1.70}$
		&$2 .20^{+1.10}_{-1.70}$
		&$1 .52^{+0.95}_{-1.63}$
		&&$1 .02^{+0.50}_{-0.78}$
		&$1 .21^{+0.51}_{-0.76}$
		&$0 .77^{+0.47}_{-0.80}$\\
		
		$\sum m_{\nu}$ [eV]         & $<0.302$
		& $<0.317$
		& $<0.282$
		&& $<0.268$
		& $<0.288$
		& $<0.250$\\
		
		$\Omega_{\rm m}$      & $0 .3094\pm0.0082$
		&$0 .3106^{+0.0083}_{-0.0082}$
		&$0 .3080^{+0.0081}_{-0.0089}$
		&&$0 .3066\pm0.0072$
		&$0 .3078\pm0.0072$
		&$0 .3056\pm0.0074$\\
		
		$H_0$ [km/s/Mpc]        & $68 .31\pm0.82$
		&$68 .27^{+0.83}_{-0.82}$
		&$68 .33^{+0.82}_{-0.81}$
		&&$68 .54\pm0.74$
		&$68 .52\pm0.75$
		&$68 .53^{+0.75}_{-0.74}$\\

		$S_8$       & $0 .825\pm0.012$
		&$0 .823\pm0.012$
		&$0 .827\pm0.012$
		&& $0 .824\pm0.012$
		&$0 .822\pm0.011$
		&$0 .826^{+0.013}_{-0.012}$\\

		\hline
		$\chi^{2}_{\rm min}$    &$3822.100$
		&$3822.180$
		&$3821.048$
		&&$3822.458$
		&$3823.538$
		&$3821.284$
		\\
		\hline\hline
	\end{tabular}
\end{table*}

\begin{table*}\tiny
	\caption{The fitting values for the Sin+$\sum m_{\nu}$ and MSin+$\sum m_{\nu}$ models considered mass hierarchy cases of
		NH, IH, and DH.}
	\label{tab4}
	\small
	\setlength\tabcolsep{2.8pt}
	\renewcommand{\arraystretch}{1.5}
	\centering
	\begin{tabular}{cccccccccccc}
		\\
		\hline\hline &\multicolumn{3}{c}{Sin} &&\multicolumn{3}{c}{MSin} \\
		\cline{2-4}\cline{6-8}
		Parameter  &NH &IH &DH &&NH &IH  &DH  \\ \hline
		
		$w_{0}$                & $-0.956^{+0.063}_{-0.070}$
		&$-0.952^{+0.065}_{-0.066}$
		&$-0.962\pm0.063$
		&& $-1$
		& $-1$
		& $-1$\\

		$w_{1}$       & $0 .80^{+0.37}_{-0.70}$
		&$0 .91^{+0.41}_{-0.69}$
		&$0 .66^{+0.34}_{-0.69}$
		&& $0 .49^{+0.21}_{-0.38}$
		&$0 .57^{+0.22}_{-0.38}$
		&$0 .37^{+0.19}_{-0.39}$\\
		
		$\sum m_{\nu}$ [eV]         & $<0.327$
		& $<0.336$
		& $<0.311$
		&& $<0.298$
		& $<0.318$
		& $<0.277$\\
		
		$\Omega_{\rm m}$       & $0 .3097^{+0.0083}_{-0.0090}$
		&$0 .3106^{+0.0082}_{-0.0083}$
		&$0 .3081\pm0.0084$
		&& $0 .3069\pm0.0072$
		&$0 .3079\pm0.0072$
		&$0 .3058\pm0.0073$\\
		
		$H_0$ [km/s/Mpc]        & $68 .33^{+0.83}_{-0.84}$
		&$68 .32^{+0.84}_{-0.83}$
		&$68 .37\pm0.82$
		&& $68 .57^{+0.72}_{-0.73}$
		&$68 .55\pm0.73$
		&$68 .56^{+0.74}_{-0.73}$\\

		$S_8$       & $0 .825\pm0.012$
		&$0 .823\pm0.012$
		&$0 .826\pm0.012$
		&& $0 .824\pm0.012$
		&$0 .822\pm0.012$
		&$0 .826\pm0.012$\\

		\hline
		$\chi^{2}_{\rm min}$    &$3822.408$
		&$3823.456$
		&$3821.080$
		&&$3822.876$
		&$3823.574$
		&$3821.224$
		\\
		\hline\hline
	\end{tabular}
\end{table*}

\begin{figure*}[ht!]
	\begin{center}
		\includegraphics[width=5.5cm]{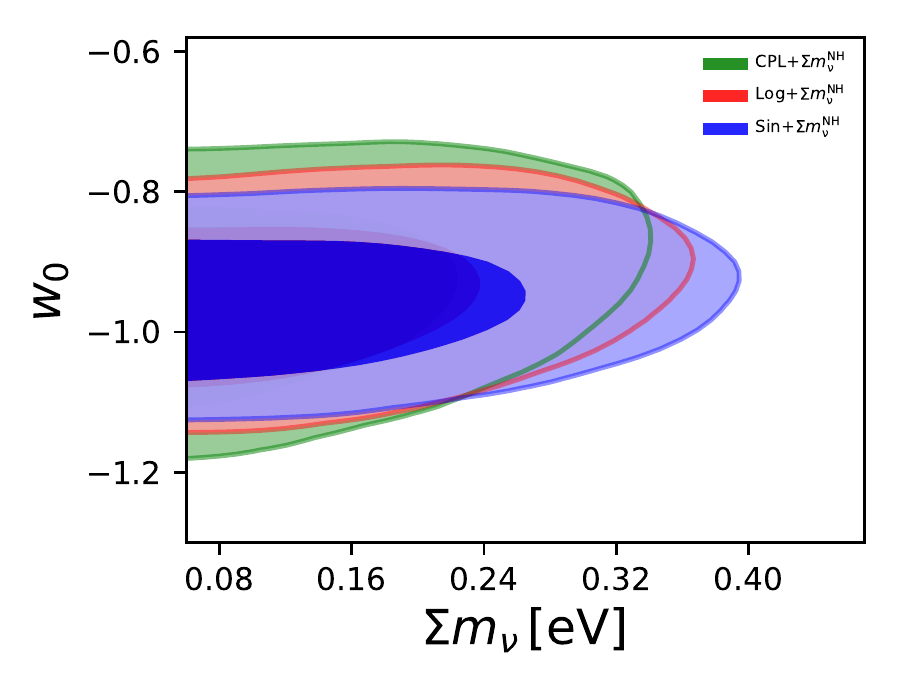}
		\includegraphics[width=5.5cm]{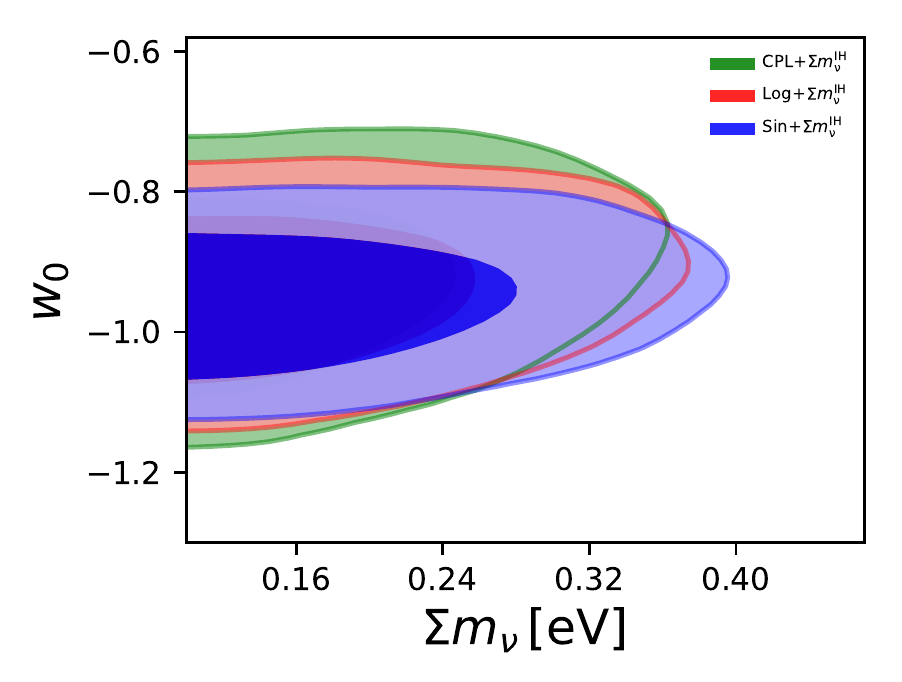}
		\includegraphics[width=5.5cm]{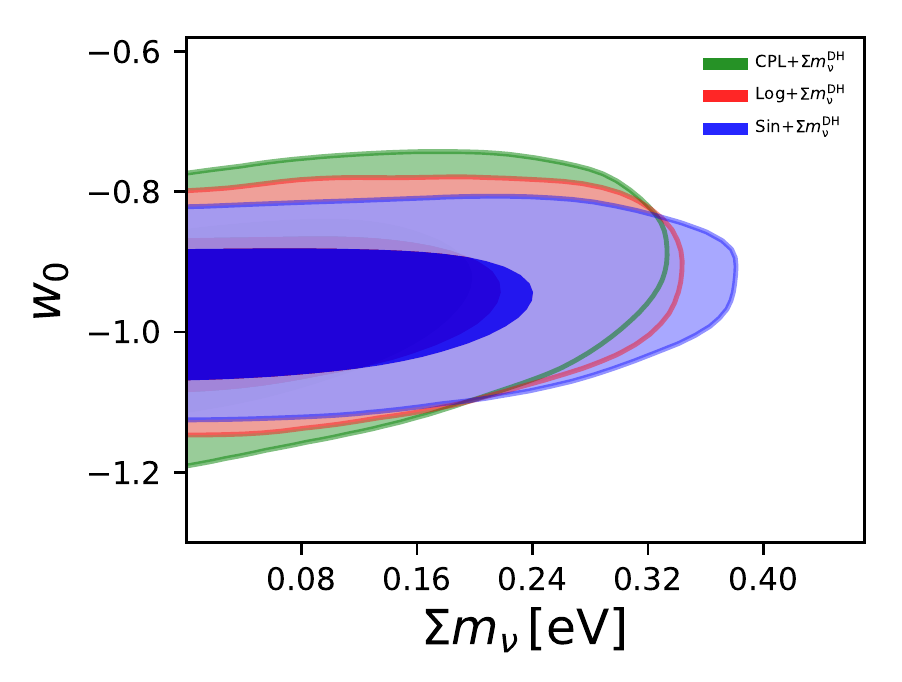}
	\end{center}
	\caption{Two-dimensional marginalized contours (68.3\% and 95.4\% confidence level) in the $\sum m_{\nu}$–$w_{0}$ plane of the CPL, Log, and Sin models considered mass hierarchy cases of NH, IH, and DH.}
	\label{f3}
\end{figure*}

Considering the same neutrino mass ordering, the fitting value of $\sum m_{\nu}$ is smallest in the CPL model and largest in the Sin model, confirming that the fitting values of $\sum m_{\nu}$ can be changed by modifying the $w(z)$ forms. In Figure~\ref{f3}, we provide two-dimensional marginalized contours (68.3\% and 95.4\% confidence level) in the $\sum m_{\nu}$–$w_{0}$ plane of the CPL, Log, and Sin models, considered mass hierarchy cases of NH, IH, and DH. In the three two-parametrization models, $\sum m_{\nu}$ is positively correlated $w_{0}$, which ensures the same observed acoustic peak scale in the cosmological fit using the Planck data. When we compare the constraint results of $\sum m_{\nu}$ for the three different cases of neutrino mass orderings, we find that the smallest value of $\sum m_{\nu}$ is obtained in the DH case, and the largest value of $\sum m_{\nu}$ corresponds to the IH case, which mean that considering the mass hierarchy can also affect the fitting values of $\sum m_{\nu}$.

In the CPL+$\sum m_{\nu}$ model, we obtain $\chi^{2}_{\rm min}=3822.102$ for the NH case, $\chi^{2}_{\rm min}=3822.516$ for the IH case, and $\chi^{2}_{\rm min}=3821.168$ for the DH case (see Table~\ref{tab2}). In the Log+$\sum m_{\nu}$ model, we have $\chi^{2}_{\rm min}=3822.100$ for the NH case, $\chi^{2}_{\rm min}=3822.180$ eV for the IH case, and $\chi^{2}_{\rm min}=3821.048$ for the DH case (see Table~\ref{tab3}). In the Sin+$\sum m_{\nu}$ model, the constraint results become $\chi^{2}_{\rm min}=3822.408$ for the NH case, $\chi^{2}_{\rm min}=3823.456$ for the IH case, and $\chi^{2}_{\rm min}=3821.080$ for the DH case (see Table~\ref{tab4}). Obviously, the small difference of the $\chi^{2}_{\rm min}$ values among the three mass hierarchies only stems from the different prior ranges of the patrameter $\sum m_{\nu}$, which does not help to distinguish the neutrino mass orderings. 

We also discuss the constraints of $\sum m_{\nu}$ in the MCPL model, the MLog model, and the MSin model, in which $w(z)$ is parameterized with a single free parameter $w_{1}$. In the MCPL+$\sum m_{\nu}$ model, we obtain $\sum m_{\nu}<0.250$ eV for the NH case, $\sum m_{\nu}<0.276$ eV for the IH case, and $\sum m_{\nu}<0.228$ eV for the DH case (see Table~\ref{tab2}). In the MLog+$\sum m_{\nu}$ model, we have $\sum m_{\nu}<0.268$ eV for the NH case, $\sum m_{\nu}<0.288$ eV for the IH case, and $\sum m_{\nu}<0.250$ eV for the DH case (see Table~\ref{tab3}). In the MSin+$\sum m_{\nu}$ model, the constraint results become $\sum m_{\nu}<0.298$ eV for the NH case, $\sum m_{\nu}<0.318$ eV for the IH case, and $\sum m_{\nu}<0.277$ eV for the DH case (see Table~\ref{tab4}). Not surprisingly, the constraint results of $\sum m_{\nu}$ are largest in the MSin model and smallest in the MCPL model. 

\begin{figure*}[ht!]
	\begin{center}
		\includegraphics[width=5.5cm]{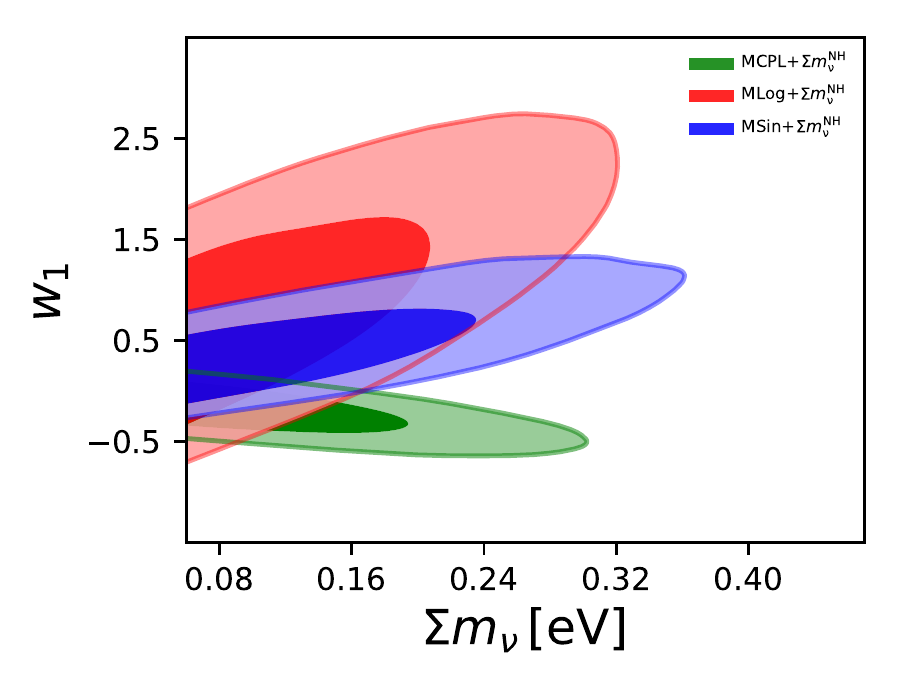}
		\includegraphics[width=5.5cm]{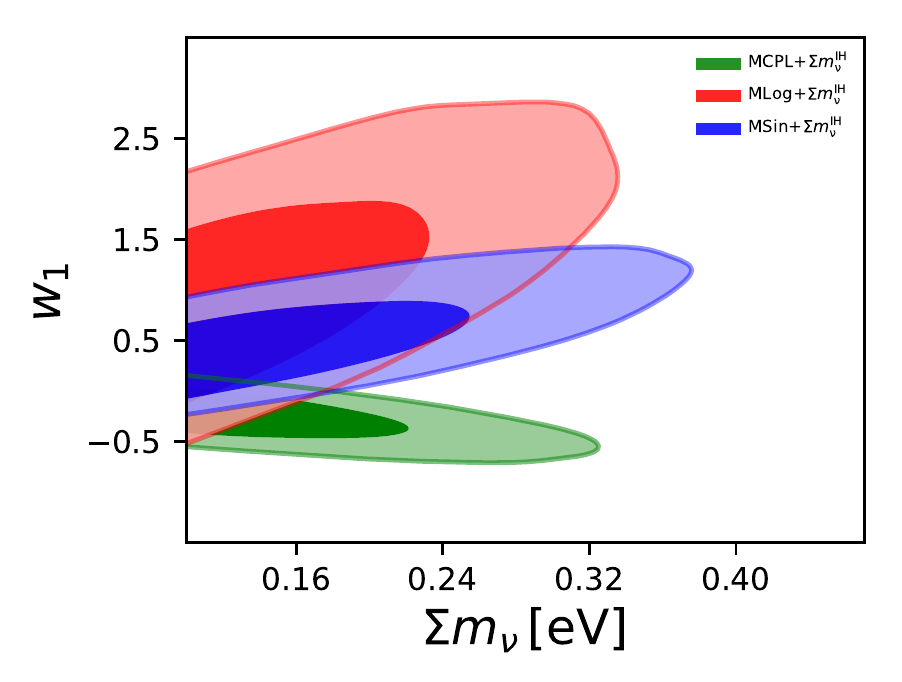}
		\includegraphics[width=5.5cm]{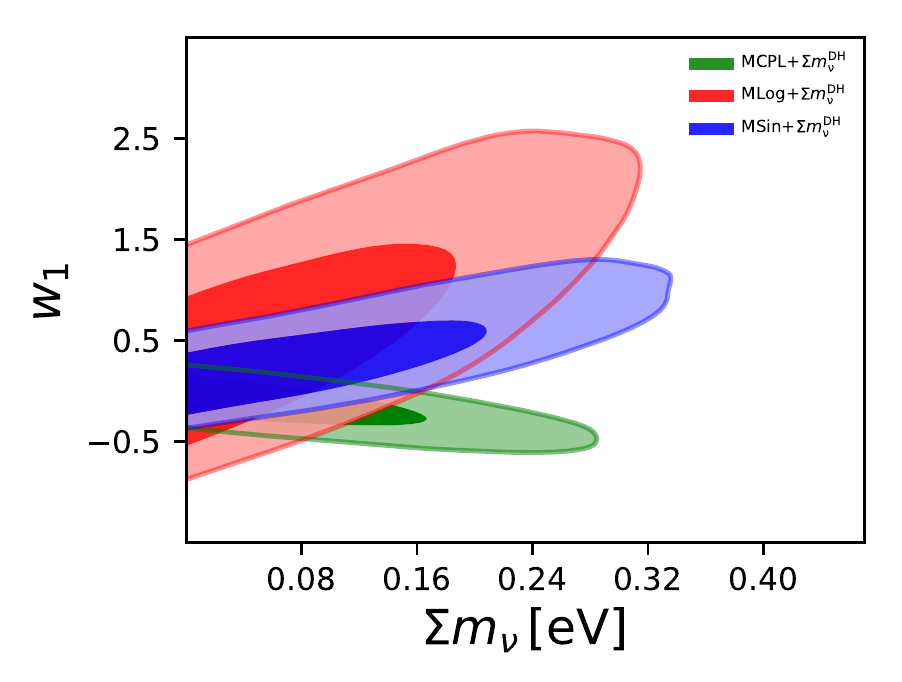}
	\end{center}
	\caption{Two-dimensional marginalized contours (68.3\% and 95.4\% confidence level) in the $\sum m_{\nu}$–$w_{1}$ plane of the MCPL, MLog, and MSin models considered mass hierarchy cases of NH, IH, and DH.}
	\label{f4}
\end{figure*}

Furthermore, comparing constraint results of $\sum m_{\nu}$ with those derived from the two-parametrization models, we find that the values of $\sum m_{\nu}$ are smaller in these one-parametrization models, indicating that a model with less parameters tends to provide a smaller fitting value of $\sum m_{\nu}$. The two-dimensional marginalized contours in the $\sum m_{\nu}$–$w_{1}$ plane are shown in Figure~\ref{f4}. We see that $\sum m_{\nu}$ is positively correlated with $w_{1}$ in the MCPL and MLog models, but is anti-correlated with $w_{1}$ in the MSin model. The different degeneracies between them ensure that  the ratio of the sound horizon and angular diameter distance remains nearly constant.

\section{Conclusion}\label{sec:4}

In this paper, we revisit the constraints on three dynamical dark energy models that are parameterized by two free parameters, $w_{0}$ and $w_{1}$. They correspond to the CPL parametrization, the logarithm parametrization, and the oscillating parametrization. We employ current cosmological observations including the CMB data, the BAO data, and the SNe data. We obtain almost identical $\chi^{2}_{\rm min}$ values ($\Delta \chi^{2}_{\rm min} \leq 0.064$) in the three models, meaning that the Log model and the Sin model can behave as the same as the conventional CPL model in the fit to the CMB+BAO+SNe data. But the advantage of the logarithm parametrization and the oscillating parametrization over the CPL model is that they can overcome the future divergency problem, and successfully probe the dynamics of dark energy in all the evolution stages of the universe. Furthermore, compared to the base $\Lambda$CDM model, we find that the two novel parametrizations with $\Delta \chi^{2}_{\rm min} \leq -3$ are substantially supported by the CMB+BAO+SNe data. 

We investigate the constraints on the total neutrino mass $\sum m_{\nu}$ in these dynamical dark energy. Meanwhile, we consider the NH case, the IH case, and the DH case of three-generation neutrino mass. We find that the smallest fitting value of $\sum m_{\nu}$ is obtained in the DH case, and the largest value of $\sum m_{\nu}$ corresponds to the IH case in thses models. For example, we have $\sum m_{\nu}<0.302$ eV for the NH case, $\sum m_{\nu}<0.317$ eV for the IH case, and $\sum m_{\nu}<0.282$ eV for the DH case, in the Log model. Such results tell us that the different neutrino mass hierarchies affect the constraint results of $\sum m_{\nu}$. However, our constraints results does not provide more evidence for determining the neutrino mass orderings, owing to the larger fitting values of $\sum m_{\nu}$ and the similar values of $\chi^{2}_{\rm min}$ obtained for different neutrino mass hierarchies. 

For the models with different parametrizations of dark energy, we find that the values of $\sum m_{\nu}$ in the Log and Sin models are larger than those derived from the CPL model. For example, we obtain $\sum m_{\nu}<0.285$ eV for the CPL model, $\sum m_{\nu}<0.302$ eV for the Log model, and $\sum m_{\nu}<0.327$ eV for the Sin model, in the NH case. For the IH and DH cases, the conclusion is the same. Thus our results confirm the conclusion that the dark energy properties could indeed significantly change the fitting results of $\sum m_{\nu}$. In addition, we discuss the case that $w_{0}=-1$ is fixed in the three dynamical dark energy models. The conclusions remain the same as those derived in the investigation of the constraints on the CPL model, the Log model, and the Sin model.

As a summary, our conclusions in this work are (i) The logarithm parametrization and the oscillating parametrization for dark energy are substantially supported by current observational data. (ii) The two parametrizations for dark energy can increase the fitting value of $\sum m_{\nu}$. (iii) The different neutrino mass hierarchies can affect the constraint results of $\sum m_{\nu}$. But a special mass hierarchy (NH or IH) is not determined in the two parametrizations.

\begin{acknowledgments}

This work is supported by the National Natural Science Foundation of China (Grant No. 12103038).

\end{acknowledgments}

\end{document}